\documentclass{jkas}
\def\year{2021} 
\def\month{xxx} 
\def\volume{00} 
\def\issue{0} 
\def\beginpage{1} 
\def\endpage{8} 
\def\received{June 26, 2021} 
\def\accepted{March 15, 2023} 
\date{Received \received; accepted \accepted}

\title{A Search for Exoplanets around Northern Circumpolar Stars\\
VII. Detection of planetary companion orbiting the largest host star HD~18438}

\author[1, 2]{Byeong-Cheol Lee}
\author[3]{Jae-Rim Koo}
\author[1]{Gwanghui Jeong}
\author[4]{Myeong-Gu Park}
\author[1]{Inwoo Han}
\author[1, 2]{Yeon-Ho Choi}

\affil[1]{Korea Astronomy and Space Science Institute, 776, Daedeokdae-Ro, Youseong-Gu, Daejeon 34055, Korea; \email{tlotv1@gmail.com; bclee@kasi.re.kr; iwhan@kasi.re.kr}}
\affil[2]{Astronomy and Space Science Major, University of Science and Technology, Gajeong-ro Yuseong-gu, Daejeon 305-333, Korea}
\affil[3]{Kongju National University, Gongjudaehak-ro 56, Gongju-si, Chungcheongnam-do 32588, Korea; \email{kkoojr@gmail.com}}
\affil[4]{Department of Astronomy and Atmospheric Sciences, Kyungpook National University, Daegu 702-701, Korea; \email{mgp@knu.ac.kr}}
\def\corrauthor{
Byeong-Cheol Lee
}

\def\runningauthor{
Byeong-Cheol Lee et al.
}

\def\runningtitle{
A Search for Exoplanets around Northern Circumpolar Stars
}

\def\keywords{star: individual: HD 18438 ---  techniques: radial velocities
}

\def\abstracttext{
We have been conducting a exoplanet search survey using Bohyunsan Observatory Echelle Spectrograph (BOES) for the last 18 years.
We present the detection of exoplanet candidate in orbit around HD 18438 from high-precision radial velocity (RV) mesurements. The target was already reported in 2018 (Bang et al. 2018). They conclude that the RV variations with a period of 719 days are likely to be caused by the pulsations because the Lomb-Scargle periodogram of HIPPARCOS photometric and H$_{\alpha}$ EW variations for HD 18438 show peaks with periods close to that of RV variations and there were no correlations between bisectors and RV measurements. However, the data were not sufficient to reach a firm conclusion.
We obtained more RV data for four years.
The longer time baseline yields a more accurate determination with a revised period of  803 $\pm$ 5 days and the planetary origin of RV variations with a minimum planetary companion mass of 21 $\pm$ 1 $M_{\rm{Jup}}$.
Our current estimate of the stellar parameters for HD~18438 makes it currently the largest star with a planetary companion.
}

\begin{document}
\jkashead 

\section{Introduction}
Until now, more than 4700 planets have been found in various stellar luminosity types. Specially, just $\sim$ 100 planets were detected in giant stars.
Giants are likely to show more complex RV variations because of the direct effect of various surface processes on the line profiles:
stellar pulsations, chromospheric activities, spots, and large convection cells.

In 2010, we started the Search for Exoplanet around Northern circumpolar Stars (SENS; \citealt{Lee2015}) at  Bohyunsan Astronomical Observatory (BOAO).
The main goal of the SENS is to observe stars that are accessible year round in order to have better sampling for our targets and thus
increase the planet detection efficiency.
From the SENS survey, we detected twenty planetary companions (\citealt{Lee2015}; \citealt{Lee2017}; \citealt{2018JKAS...51...17B}; \citealt{Jeo2018}; \citealt{Lee2020}) and several periodic RV variations, probably, due to processes other than orbital motions around  G-, K-, and M-giant stars.
Lack of knowledge about planet formation and evolution makes RV surveys of giant stars an important endeavor.

Bang et al. (2018; hereafter, TYB18) obtained precise RV measurements for HD~18438 using Bohyunsan Observatory Echelle Spectrograph (BOES; \citealt{Kim2007}) and
found long-period radial velocity (RV) variations with period of 719.0 days.
They concluded that the observed RV variations are likely to be caused by the pulsations because the Lomb-Scargle periodograms of HIPPARCOS photometric and H$_{\alpha}$ EW variations for HD~18438 show apparently non-negligible peaks at periods close to that of RV variations.
However, at the time, a definite conclusion could not be reached due to the lack of data, uncertainty in verification, and the amplitude of the RV variation was at odds with typical Long Secondary Periods in giants.

In this paper, we present more robust detection of low-amplitude and long-period RV variations in HD~18438 with additional RV data obtained for four more years, possibly caused by a planetary companion.
What the criteria for low amplitude and long-term RV variation mean is related to giants, generally based on periods below a km\,s$^{-1}$  and over hundreds of days.
In Sect. 2, we describe the observations and data reduction. In Sect. 3, the stellar characteristics of the host stars are derived. The orbital solutions of RV variation measurements are presented in Sect. 4. In Sect. 5, possible origins of RV variations were explained.  Finally, we summarize the study in Sect. 6.

\section{OBSERVATIONS AND REDUCTION}
\begin{table*}[t]
\caption{RV measurements for HD 18438 from November 2010 to April 2021. \label{tab:table1}}
\centering
\begin{tabular}{crccrccrc}
\toprule
JD-2450000 & $\Delta RV$ & $\pm \sigma$ & JD-2450000 & $\Delta RV$ & $\pm \sigma$ & JD-2450000 & $\Delta RV$ & $\pm \sigma$ \\
(Days) & (m\,s$^{-1}$) & (m\,s$^{-1}$) & (Days) & (m\,s$^{-1}$) & (m\,s$^{-1}$) & (Days) & (m\,s$^{-1}$) & (m\,s$^{-1}$) \\
\midrule
5529.076993 &    $-$127.0   &     9.4 &   7378.110490  &    334.3  &      9.7  &  8863.099371  &   $-$93.3   &    11.3    \\
5842.233812 &    $-$113.4   &     9.4 &   7401.925174  &    203.3  &     10.8  &  8932.000976  &     25.0    &   14.0     \\
5933.100505 &       142.3   &    10.9 &   7414.948654  &    392.4  &      9.6  &  8933.018910  &     34.7    &   12.5     \\
5962.981233 &        36.6   &    10.8 &   7423.943333  &    190.7  &     10.1  &  8942.949838  &     79.8    &   10.1     \\
6259.127756 &    $-$321.8   &     9.7 &   7468.938303  &    271.1  &      9.6  &  8944.996838  &    151.4    &   12.2     \\
6287.033915 &    $-$361.1   &    12.9 &   7490.997465  &    275.6  &     25.3  &  8969.977387  &    263.2    &   12.5     \\
6288.154197 &    $-$373.9   &    10.1 &   7491.008796  &    232.1  &     11.6  &  9134.958971  &    153.0    &   11.1     \\
6347.036601 &    $-$306.9   &     9.2 &   7527.980176  &     57.1  &     11.7  &  9134.971889  &    150.8    &   12.1     \\
6551.230025 &       173.5   &    12.4 &   7528.978922  &    109.0  &     11.3  &  9149.294967  &     75.3    &    9.4     \\
6578.267935 &        65.4   &    10.8 &   7529.966975  &    218.2  &     11.1  &  9149.306854  &     80.2    &    9.2     \\
6616.998903 &       148.5   &     8.1 &   7530.976243  &    288.2  &     11.8  &  9150.025131  &     61.7    &   11.8     \\
6714.059556 &       307.1   &    13.3 &   7672.942073  &   $-$60.9 &      14.0 &   9150.994632 &      38.6   &     8.3    \\
6739.952656 &       200.3   &    11.2 &   7703.900322  &  $-$216.9 &       9.2 &   9161.069290 &     114.7   &     9.4    \\
6808.051845 &        97.1   &    13.0 &   7705.102409  &  $-$184.4 &       8.0 &   9161.083075 &     120.0   &    10.9    \\
6808.061359 &        93.3   &    15.0 &   7757.052180  &  $-$328.9 &      11.7 &   9162.182363 &     136.6   &    10.7    \\
6808.068234 &       115.9   &    17.4 &   7758.076241  &  $-$359.1 &       9.2 &   9216.967993 &     $-$1.1  &     11.2   \\
6922.125007 &    $-$175.0   &    13.3 &   7854.998117  &   -400.5  &     11.0  &  9218.049565  &      38.4   &    12.2    \\
6964.052704 &    $-$393.2   &    11.9 &   8109.217336  &    115.3  &     10.6  &  9295.964472  &    $-$244.2 &      10.4  \\
6965.262046 &    $-$468.8   &     9.3 &   8148.023617  &    310.7  &     10.9  &  9298.964320  &    $-$179.1 &      10.7  \\
7094.000813 &    $-$236.0   &    14.3 &   8515.925612  &  $-$224.5 &      10.9 &   9302.005836 &    $-$167.3 &      16.7  \\
7298.015438 &      $-$6.7   &    11.7 &   8830.219372  &  $-$129.4 &      10.1 &   9330.961200 &    $-$398.9 &      12.9  \\
7330.115826 &        73.4   &     8.7 &   8862.014128  &   $-$99.9 &       8.7 &               &             &            \\
 \bottomrule
 \end{tabular}
 \end{table*}

The fiber-fed, high-resolution ($\emph{R}$ = 45 000) BOES installed at the 1.8-m telescope of BOAO \citep{Kim2007}, Korea was used for all RV measurements.
An iodine absorption (I$_{2}$) cell was used with a wavelength region of 4900$-$5900 {\AA}. The average signal-to-noise (S/N) for the I$_{2}$ region was
about 150 at typical exposure times ranging from 10 to 20 minutes.

In 2018, TYB18 reported the first result from seven years observations (from November 2010 to April 2017). We obtained additional spectra for the following four years (from December 2017 to April 2021).
The IRAF software package was used for the basic reduction of spectra and precise RV measurements utilizing the I$_2$ cell were carried out with the RVI2CELL \citep{Han2007} code, which is based on the method by \citet{But1996} and \citet{Val1995}.
To demonstrate a long-term RV stability, the RV standard star $\tau$ Ceti was observed.
RVs measured by the BOES are constant with an rms scatter of $\sim$ 7 m s$^{-1}$ \citep{Lee2013}.
In this paper, all data including the BTY18 paper were analyzed.
The resultant RV measurements are listed in Table~\ref{tab:table1}.

\section{STELLAR CHARACTERISTICS}
%
\begin{table*}[t]
\caption {Stellar parameters for HD 18438 and TYC 4516-2148-1 \label{tab:table2}}
\centering
\begin{tabular}{lcccc}
\toprule
Parameter & Unit & HD 18438 & TYC 4516-2148-1 & Reference \\
\midrule
    Spectral type      &     & M2.5 III &  & \citet{2008ApJS..176..216A}  \\
    $\textit{$m_{v}$}$ &[mag]& 5.49  &  & \emph{HIPPARCOS} (\citealt{ESA1997})  \\
                       &[mag]&  & 9.08 & \citet{Fab2002}  \\
    $\textit{B-V}$     &[mag]& 1.569 $\pm$ 0.015 & &  van Leeuwen (2007) \\
    age                &[Gyr]& 5.5 $\pm$ 2.4 & & This work   \\
    $\pi$              &[mas]& 4.443 $\pm$ 0.171 & 4.223 $\pm$ 0.035 &Gaia collaboration (2018)  \\
    $T_{\rm{eff}}$     &[K] & 3860 $\pm$ 100 &                          &    Gaia collaboration (2018) \\
                        &[K] &               & 6164 $\pm$ 211  &  \citet{2019AJ....158..138S} \\
    $\rm{[Fe/H]}$      &[dex]& -- 0.4 $\pm$ 0.1 & &  \citet{2018JKAS...51...17B}  \\
    log $\it g$        &[cgs]& 0.9  $\pm$ 0.1 &                & This work  \\
                        &[cgs]&               & 3.7  $\pm$ 0.1 & \citet{2019AJ....158..138S}  \\
    $v_{\rm{micro}}$   &[km s$^{-1}$]& 2.7 $\pm$ 0.4 & &    \citet{2018JKAS...51...17B}  \\
    $\textit{$R_{\star}$}$  &[$R_{\odot}$]& 88.475 $\pm$ 4.424 &  & \citet{Ker2019}  \\
                            &[$R_{\odot}$]&                         & 2.554 $\pm$ 0.182 & \citet{2019AJ....158..138S}\\
    $\textit{$M_{\star}$}$  &[$M_{\odot}$]&  1.84 $\pm$ 0.09 & & \citet{Ker2019} \\
                            &[$M_{\odot}$]&                    & 1.174 $\pm$ 0.184 & \citet{2019AJ....158..138S}  \\
    $\textit{$L_{\star}$}$  &[$L_{\odot}$] & 929 $\pm$ 41 & 7.558 $\pm$ 0.095  & Gaia collaboration (2018) \\
    $v_{\rm{rot}}$ sin($i$)  &[km s$^{-1}$] & 5.5 $\pm$ 0.2  &  & \citet{2018JKAS...51...17B} \\
    $P_{\rm{rot}}$ / sin($i$)  &[days] &  637  &  &   This work \\
\bottomrule
\end{tabular}
\end{table*}

\citet{Fab2002} reported that giant HD~18438 is likely to be a binary system with an orbital separation  of $\sim$ 1100 AU, including the main sequence star (TYC 4516-2148-1).  Although the proper motion and parallax are almost the same and their masses are almost the same, it is still difficult to understand how they are related. Interestingly, however, a secondary mass (normalized to 1 AU) of 21.72 $M_{\rm{Jup}}$ $^{+20.62} _{-19.84}$ was estimated around the giant HD~18438.

The main parameters for HD 18438 was acquired from the Gaia database (\citet{Gai2018}; \citet{Ker2019}; \citet{2019AJ....158..138S}) and the \emph{HIPPARCOS} catalog (\citealt{ESA1997}).
Table~\ref{tab:table2} summarizes the basic stellar parameters.

\begin{figure}[t]
\centering
\includegraphics[width=9cm]{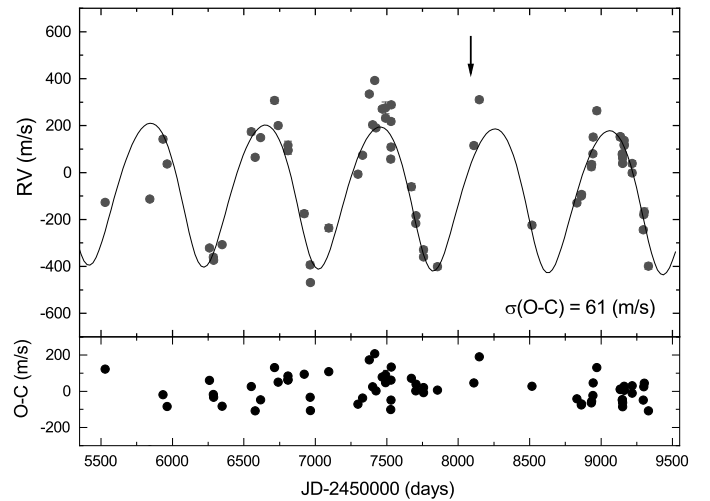}
\caption{RV measurements (top panel) and the residuals (bottom panel) for HD~18438 from November 2010 to June 2021.
The solid line is the orbital motion with a period of 803 days. The arrow marks April 2017, when we restarted to obtain more RV data.}
    \label{fig:orbit}
\end{figure}

\begin{figure}
\centering
\includegraphics[width=8cm]{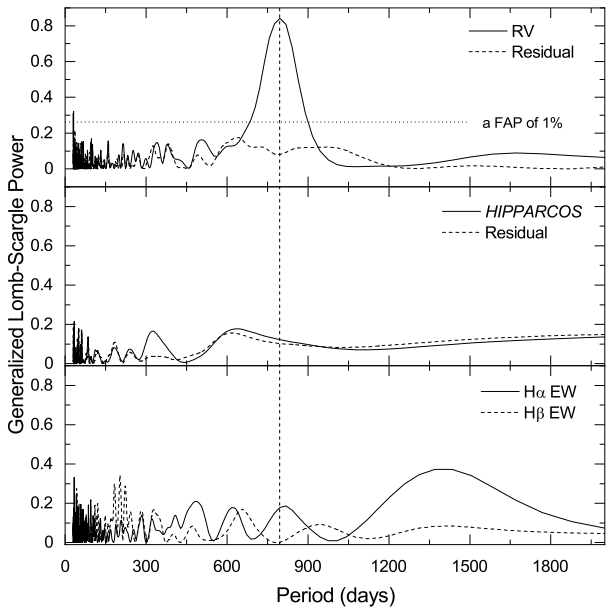}
    \caption{The GLS periodograms of the RV measurements, the \emph{HIPPARCOS} photometry data, and the EW variations of the hydrogen lines (top to bottom panel) from 2010 to 2021. The vertical dashed line indicates the location of the period of 803 days. The solid line in the top panel is the GLS periodogram of the RV measurements for whole 11 years. The dashed line is the periodogram of the residual. The horizontal dotted line shows a FAP threshold of 1 $\times 10^{-2}$ (1\%).}
       \label{fig:power}
\end{figure}

   \begin{figure}[t]
   \centering
   \includegraphics[width=8cm]{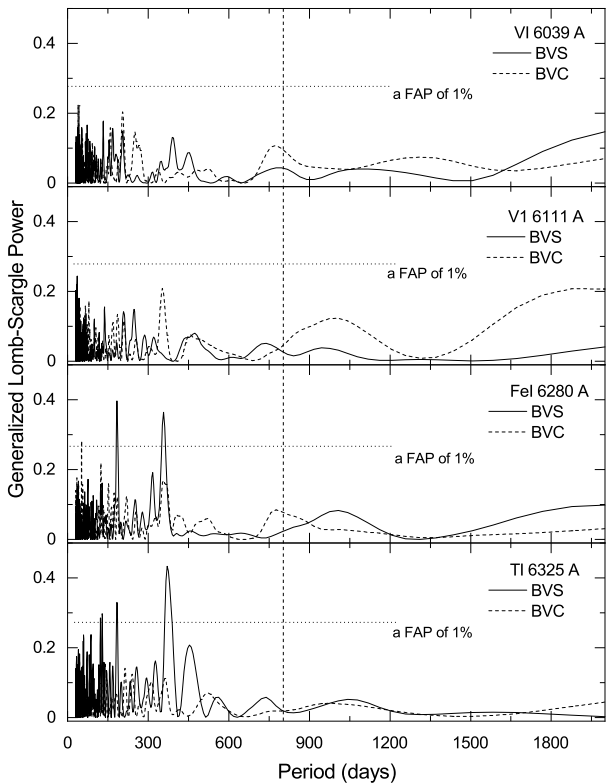}
      \caption{The GLS periodograms of the RV period and the four kinds of line bisectors for HD~18438.
      The horizontal dotted line shows a FAP threshold of 1 $\times 10^{-2}$ (1\%) and the vertical dashed line indicates the location of the period of 803 days.}
        \label{fig:bisector}
   \end{figure}

 \begin{figure}[t]
   \centering
   \includegraphics[width=8cm]{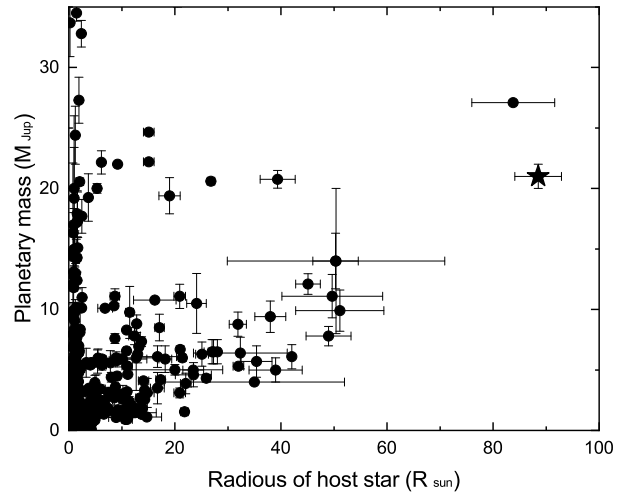}
      \caption{Distribution of stellar radii vs. minimum masses of planetary companions as of July 2021. Pentagram denotes the planetary companion around HD 18438.}
        \label{fig:sat}
   \end{figure}

\section{ORBITAL SOLUTIONS OF RADIAL VELOCITY VARIATIONS}
%
In order to find the periodicity in the RV variations, we used the Generalized Lomb-Scargle periodogram (GLS; Zechmeister \& Kürster 2009).
Compared to the 'classical' Lomb-Scargle periodogram,
it provides more accurate frequencies while less susceptible to aliasing.
We combined all RV measurements by BOES into a single RV set, including 27 RVs taken over the last four years.

The resulting  RV curve as a function of the time for HD 18438 is shown in Figure ~\ref{fig:orbit}.
The arrow in the figure denotes the time when we started to obtain additionally spectra after previously observed last point by TYB18.
All the data were analyzed, yielding a bit different orbital solutions. Period changed from 719 to 803 days, semi-amplitude from 205 to 305 m\,s$^{-1}$, and an rms scatter ($\sigma$(O-C)) from 55 to 61 m\,s$^{-1}$.
TYB18 reported two significant peaks at values of 719 days and $\sim$ 800 days in calculated RV periodogram in their Figure 5. They selected a relatively higher peak the last (719 days) as the main period. However, the whole RV set displays a unique peak with much increased confidence (top panel in Fig.~\ref{fig:power}).
Assuming a stellar mass of 1.84 $\pm$ 0.09~$M_{\odot}$, minimum mass of the planetary companion is 21 $\pm$ 1 $M_{\rm{Jup}}$ at a distance of 2.1 $\pm$ 0.1 AU from the host.
A false alarm may appear in period analysis techniques when a period is incorrectly found where none exists in reality.
Experimentally, FAPs below 0.01 (1\%) mostly indicate a very secure period.
Keplerian orbital elements are listed in Table~ \ref{tab:table3}.

\begin{table*}
\caption{Orbital parameters for HD 18438 b \label{tab:table3}}
\centering
\begin{tabular}{lcc}
\toprule
Parameter & Unit &  HD 18438  \\
\midrule
    Period                       &[days] & 803  $\pm$ 5       \\
    $\it T$$_{\rm{periastron}}$  &[JD]   &  2455367 $\pm$ 45  \\
    $\it{K}$                     &[m s$^{-1}$] & 305  $\pm$ 18 \\
    $\it{e}$                     &             & 0.1   $\pm$ 0.1  \\
    $\omega$                     &[deg]      & 167   $\pm$ 24  \\
    slope           &[m s$^{-1}$ day$^{-1}$] & $-$8.5 $\times$ 10$^{- 6}$ \\
    Nobs            &--          & 65                           \\
    $\sigma$ (O-C)               &[m s$^{-1}$]  & 61           \\
\hline
    $m$ sin($i$)                 &[$M_{\rm{Jup}}$]& 21 $\pm$ 1.0   \\
    $\it{a}$                     &[AU]            & 2.1 $\pm$ 0.1 \\
\bottomrule
\end{tabular}
\end{table*}

\section{ORIGIN OF RADIAL VELOCITY VARIATIONS}

Low-amplitude and long-term periodic RV variations in evolved stars may be caused by three kinds of phenomena: stellar pulsations, rotational modulations of surface features, or planetary companions.
To find the origin of the RV variations observed in HD~18438, we have performed some validations that are currently possible:
 the \emph{HIPPARCOS} photometry variations, the spectral line bisectors variations, and the stellar chromospheric activity variations.
If one of these variations is similar to RV variations, it may invalidate the planetary companion interpretation.

\subsection{PHOTOMETRIC VARIATIONS}
In order to find feasible brightness variations caused by the rotational modulation of stellar spots or pulsations,
we have analyzed the \emph{HIPPARCOS} photometric data which is the only relevant photometric data currently available.
TYB18 shows a significant peak at about 700 days, close to 719 day RV period. In addition, there is another strong peak around 350 days,
about a year, and it is half of period of a significant peak at about 700 days. However, we checked again using GLS or Lomb-Scargle periodograms and see no significant peaks at around 700 days or 350 days. In addition, no significant signal was found in the residual. (the middle panel in Fig.~\ref{fig:power}).
These may have been some calculation errors in TYB18.

\subsection{CHROMOSPHERIC ACTIVITIES}
The EW variations of Ca II H \& K, H$_{\alpha}$, H$_{\beta}$, sodium lines, and Ca~II~triplet lines are mostly
used as chromospheric activity indicators, which are sensitive to the stellar atmospheric activity.
Such activity could give a significant effect on the RV variations.
However, our data do not have enough S/N ratio to resolve the emission feature in the Ca~II~H~\&~K line cores.
Ca II triplets are also not suitable because of fringing and saturations of our CCD spectra at wavelength longer than 8000~{\AA}.

Thus, in this study we use the H$_{\alpha}$ and H$_{\beta}$ lines. It is important to avoid nearby blending lines and weak telluric lines.
We measured the H line EWs using a band pass of $\pm$ 1.0 $\rm\AA$ centered on the core of the H lines.
 The mean EWs of the H lines are measured to be 1138.4 $\pm$ 1.2 m$\rm\AA$ (H$_{\alpha}$) and 934.5 $\pm$ 0.8 m$\rm\AA$ (H$_{\beta}$).
The rms in the H$_{\alpha}$ and the H$_{\beta}$ EWs correspond to under 0.1\% variations.
The GLS periodogram of the H line EW variations are shown in the bottom panel of Fig.~\ref{fig:power}.
There is no significant power at the frequency corresponding to the period around 783 days noted in TYB18.

TYB18 was failed to avoid nearby blending lines and weak telluric lines. That is, when a 2.0 A band pass is specified, the EW change increases due to Co1 6563 $\rm\AA$ around H$_{\alpha}$ line EW and Cr 4861 $\rm\AA$ around H$_{\beta}$ line EW. In this paper, the variation is less than 0.1\% in H lines, however, with a band pass of 2.0 $\rm\AA$, the variation of $~$ 2\% is calculated (See Fig. 5 of \citealt{Lee2012}).

\subsection{LINE BISECTOR}
Stellar rotational modulations of surface inhomogeneities can create variable asymmetries in the spectral line profiles (\citealt{Que2001}):
the RV difference of the central values at high and low flux levels of the line profiles (BVS: bisector velocity span) and the difference of the velocity span of the upper half and lower half of the bisectors (BVC: the velocity curvature)(\citealt{Hat2005}; \citealt{Lee2013}). In order to calculate bisectors, we selected four unblended strong lines of V I 6039.7, V I 6111.6,  Fe I 6280.6 {\AA} and Ti I 6325.2 {\AA} that are located beyond the I$_{2}$ absorption region.
To avoid the spectral core and wings, the BVS of the profile between two different flux levels at 0.8 and 0.4 of the central depth was used as the span points in the estimation.
Figures~\ref{fig:bisector} show the GLS periodograms of the each BVS and BVC for HD~18438. None of them show any meaningful periodic variations.
Although peaks are observed around $\sim$ 200 and $\sim$ 360 days, they are not found in photometric data or analysis of chromospheric activities, making it difficult to estimate a origin. Thus, long-term observations may be helpful to determine the exact cause of star with active surface activity.

\section{DISCUSSION}
TYB18 found a long-period RV variation of 719 days in M2.5 giant HD 18438 and checked the \emph{HIPPARCOS} photometry data, chromospheric activity, and bisectors in order to identify its origin.
Of these, H$_{\alpha}$ EW  and the \emph{HIPPARCOS} photometric variations are close to that of RV variations of 719 days. They, thus, conclude that the RV variation may be caused by stellar pulsations.

We had four more years of observation since and revised the orbit solution for the whole data.
The analysis of the RV measurements for HD~18438 revealed a new and clean period of 803 days, which is different from 719 days of TYB18.
The H$_{\alpha}$ EW and H$_{\beta}$ EW lines were used to monitor the chromospheric activities, which, however, do not reveal any significant evidence of variation. A study on the analysis of \emph{HIPPARCOS} photometric data also does not show any meaningful variations nor the Bisectors reveal any relation with the RV measurements.

We  also checked the variation of line broadening of the CCF(cross-correlation function), and its periodicity. As mentioned in \cite{Hat2018} and \cite{Del18}, high luminosity giants such as $\gamma$ Draconis and NGC 4349 No. 127 may have RV variations due to stellar activity and/or pulsation that cannot be detected by photometry, bisectors and chromospheric lines and can be detected by CCF. However, we did not find any long-period variation on the  CCF obtained from BOES data.  We  thus  concluded that there are no low-amplitude Long-Sencondary Periods (LSP; Wood et al. 2004) related to dipole oscillating convection modes. Although  such long period pulsation may be associated with LSP, typical amplitude of RV variations in LSP is much larger that in HD 18438.

Rotation period (upper limit) of 562 days is also too short to explain the observed RV period.
All the evidence has now convinced us that there is a planetary companion.
We conclude that M2.5 giant HD~18438 host a planetary companion with a minimum mass of 21 $M_{\rm{Jup}}$.
This value is almost consistent with the secondary mass combination estimated by \citet{2019AJ....158..138S}.

The residual RV variations (61 m s$^{-1}$) after orbital fitting do not show any additional periodic signal as shown in  the top panel of Fig.~\ref{fig:power}.
The scatters are  considerably greater than the RV precision the RV standard star $\tau$~Ceti (7~m~s$^{-1}$) and the individual RV accuracies from our survey of  $\sim$ 12~m~s$^{-1}$.
The rms of K giant RV residuals has a typical value of 20 m s$^{-1}$ (Hekker et al. 2006), which increases toward later spectral type. Lee et al. (2013) have shown larger residuals from M giants, $\sim$ 39 ~m~s$^{-1}$ in M1 giant HD~208527 and $\sim$ 57 ~m~s$^{-1}$ in M2 giant HD~220074.

Figure~\ref{fig:sat} shows the distribution of currently confirmed exoplanets from The Extrasolar Planets Encyclopaedia archive\footnote{http://exoplanet.eu/} in the diagram for the radius of host stars versus the mass of planetary companions.
Of the $\sim$ 4700 host stars harboring planetary companions with known stellar radii, more 95\% are smaller than 5 $R_{\odot}$ ($\sim$~0.025 AU).
The pentagram at $R_{\star}$ = 88 $R_{\odot}$ in Fig. ~\ref{fig:sat} is HD~18438, which will be at the moment, the largest star with a planetary companion.
The rightmost star is HD 81817, which is the largest star and happens to be a hybrid star, with a brown dwarf candidate (Bang et al. 2020).

The fate of planets changes dramatically as it evolves from the MS stage to the evolved stage, such as the red giant branch (RGB) and the asymptotic giant branch (AGB).
The companions may be engulfed or moved outward by the expanding star when the latter increases its radius as it evolves.
Considering its luminosity and temperature, HD~18438 is located at the AGB stage on the H–R diagram after undergoing the helium flash.
It is still unknown what happens to the companion when it interacts with the atmosphere of expanding host star.
Discovery and study of highly evolved planetary systems such as HD~18438 will help understand the late and final fates of the planetary systems.

\acknowledgments{
BCL acknowledges partial support by the KASI (Korea Astronomy and Space Science Institute) grant
2021-1-830-08 and acknowledge support by the National Research Foundation of Korea(NRF) grant funded by the Korea government(MSIT) (No.2021283200).
M.G.P. was supported by the Basic Science Research Program through the National Research Foundation of Korea (NRF) funded by the Ministry of Education (2019R1I1A3A02062242) and KASI under the R\&D program supervised by the Ministry of Science, ICT and Future Planning.
 This research made use of the SIMBAD database, operated at the CDS, Strasbourg, France.
}


\end{document}